# CLINICAL TRIAL DESIGN ENABLING ε-OPTIMAL TREATMENT RULES

Charles F. Manski
Department of Economics and Institute for Policy Research, Northwestern University

and

Aleksey Tetenov
Collegio Carlo Alberto

September 25, 2015

Abstract

Medical research has evolved conventions for choosing sample size in randomized clinical trials that rest on the theory of hypothesis testing. Bayesians have argued that trials should be designed to maximize subjective expected utility in settings of clinical interest. This perspective is compelling given a credible prior distribution on treatment response, but Bayesians have struggled to provide guidance on specification of priors. We use the frequentist statistical decision theory of Wald (1950) to study design of trials under ambiguity. We show that *ε-optimal* rules exist when trials have large enough sample size. An ε-optimal rule has expected welfare within ε of the welfare of the best treatment in every state of nature. Equivalently, it has maximum regret no larger than ε. We consider trials that draw predetermined numbers of subjects at random within groups stratified by covariates and treatments. The principal analytical findings are simple sufficient conditions on sample sizes that ensure existence of ε-optimal treatment rules when outcomes are bounded. These conditions are obtained by application of Hoeffding (1963) large deviations inequalities to evaluate the performance of empirical success rules.

1. Introduction

This paper develops criteria for choice of sample size in randomized clinical trials that, in a sense to be made precise, enable implementation of nearly optimal treatment rules in clinical practice. As background, we first review the established convention for choice of sample size, the Bayesian critique thereof, and the foundation of our work in Wald's statistical decision theory. We then introduce the concept of ε-optimal treatment rules.

1.1. Background

Medical research has evolved conventions for choosing sample size in classical randomized trials possessing perfect internal and external validity. The conventions rest on the statistical theory of hypothesis testing. To illustrate, the International Conference on Harmonisation (ICH) has provided guidance for design of trials evaluating pharmaceuticals. An ICH document states (International Conference on Harmonization, 1999, p. 1923):

> "Using the usual method for determining the appropriate sample size, the following items should be specified: a primary variable; the test statistic, the null hypothesis; the alternative ('working') hypothesis at the chosen dose(s) (embodying consideration of the treatment difference to be detected or rejected at the dose and in the subject population selected); the probability of erroneously rejecting the null hypothesis (the type I error); the probability or erroneously failing to reject the null hypothesis (the type II error), as well as the approach to dealing with treatment withdrawals and protocol violations. . . . . Conventionally the probability of type I error is set at 5% or less or as dictated by any adjustments made necessary for multiplicity considerations; the precise choice may be influenced by the prior plausibility of the hypothesis under test and the desired impact of the results. The probability of type II error is conventionally set at 10% to 20%."

Trials with samples too small to achieve conventional error probabilities are called "underpowered" and are regularly criticized as scientifically useless and medically unethical. For example, Halpern, Karlawish, and Berlin (2002) write (p. 358): "Because such studies may not adequately test the underlying hypotheses, they have been considered 'scientifically useless' and therefore unethical in their exposure of participants to the risks and burdens of human research." Ones with samples larger than needed to achieve conventional error probabilities are called "overpowered" and are sometimes criticized as unethical. For example, Altman (1980) writes (p. 1336): "A study with an overlarge sample may be deemed unethical through the unnecessary involvement of extra subjects and the correspondingly increased costs."

A common suggestion is to use the outcome of a hypothesis test to allocate a group of similar patients between a status quo treatment and an innovation. The null hypothesis is that the innovation is no better than the status quo and the alternative is that the innovation is better. Using the trial data to test the null, a health planner subsequently assigns all patients to the innovation if the null is rejected and all to the status quo treatment if the null is not rejected.

Bayesian statisticians have long criticized the use of concepts in hypothesis testing to design trials and make treatment decisions. The literature on Bayesian statistical inference rejects the frequentist foundations of hypothesis testing, arguing for superiority of the Bayesian practice of using sample data to transform a subjective prior distribution on treatment response into a subjective posterior distribution. See, for example, Spiegelhalter, Freedman, and Parmar (1994) and Spiegelhalter (2004). The literature on Bayesian statistical decision theory additionally argues that the purpose of trials is to improve medical decision making and concludes that trials should be designed to maximize subjective expected utility in decision problems of clinical interest. The sample sizes selected in such trials may differ from those motivated by testing theory. See, for example, Cheng, Su, and Berry (2003) and Berry (2004).

The Bayesian perspective on inference and decision making is compelling when a decision maker feels able to place a credible prior distribution on treatment response. However, Bayesian statisticians have long struggled to provide guidance on specification of priors and the matter continues to be controversial. See, for example, the spectrum of views expressed by the authors and discussants of Spiegelhalter, Freedman, and Parmar (1994). The controversy suggests that inability to express a credible prior is common in actual decision settings. In these circumstances, the decision maker faces a problem of choice under ambiguity (Ellsberg, 1961).

In past research, we have studied the use of trial data to make treatment decisions under ambiguity. To address the problem, we have utilized and extended the Wald (1950) development of frequentist statistical decision theory. Wald considered the broad problem of using sample data to make decisions under uncertainty. He posed the task as choice of a statistical decision function, which maps potential data into a choice among the feasible actions. He recommended ex ante evaluation of statistical decision functions as procedures, specifying how a decision maker would use whatever data may be realized. Expressing the objective as minimization of a loss function, he proposed that the decision maker evaluate a statistical decision function by the distribution of loss that it yields across realizations of the sampling process. He focused attention on mean sampling performance, which he termed risk.

Wald prescribed a three-step decision process. The first stage specifies the state space (parameter space), which indexes the set of values of unknown quantities that the decision maker deems possible. The second stage eliminates inadmissible statistical decision functions. A decision function is inadmissible (weakly dominated) if there exists another one that yields at least as good sampling performance in every possible state of nature and strictly better performance in some state. The third stage uses some criterion to choose an admissible statistical decision function. Wald considered the minimax criterion, but researchers have subsequently studied other criteria including minimax regret and minimization of a subjective mean of the risk function (Bayes risk). The minimax and minimax-regret criteria do not use a prior distribution on the state space. Hence, they are applicable to problems of decision under ambiguity.

Manski (2004, 2005, 2007), Manski and Tetenov (2007, 2014), Hirano and Porter (2009), Schlag (2006), Stoye (2009, 2012), and Tetenov (2012) have used the Wald framework to study how a health planner or similar decision maker might use sample data on treatment response to choose treatments for the members of a population. In this setting, a statistical decision function uses the data to choose a treatment allocation, so such a function has been called a *statistical treatment rule* (STR). The state space specifies the distributions of treatment response that the planner deems possible. The planner's objective has been expressed as maximization of a social welfare function that sums treatment outcomes across the population. The mean sampling performance of an STR has been called its *expected welfare*. The works cited above mainly study use of the minimax-regret criterion to choose a treatment rule, but Manski (2005) and Manski and Tetenov (2007) analyze admissibility. Manski and Tetenov (2014) extend the Wald theory by considering the quantile sampling performance of STRs rather than their mean performance.

1.2. Trials Enabling Mean ε-Optimal Treatment Rules

In this paper we turn attention to design of trials under ambiguity. Let expected welfare measure the sampling performance of an STR. Then an ideal objective for trial design is to collect data that enable implementation of a *mean optimal* rule--one whose expected welfare equals the welfare of the best treatment in every state of nature.[1]

Mean optimality is too strong a property to be achievable in general. We show, however, that *mean ε-optimal* rules do exist when trials have large enough sample size. Given a specified $\varepsilon > 0$, a mean ε-optimal rule is one whose expected welfare is within ε of the welfare of the best treatment in every state of nature. Equivalently, an ε-optimal rule has maximum regret no larger than ε. The minimum sample size needed to make ε-optimality feasible increases as ε decreases to zero.

Choosing sample size to enable existence of ε-optimal STRs provides an appealing criterion for design of trials that aim to inform clinical practice. Implementation of the idea requires specification of a value for ε. The necessity to choose an effect size of interest when designing trials already arises in conventional practice, where the trial planner must specify the alternative hypothesis to be compared with the null. A possible way to specify ε is to make it equal the *minimum clinically meaningful difference* (MCMD) in the average treatment effect comparing alternative treatments.

---

[1] The present concept of mean optimality should not be confused with various concepts of optimality in the literature on experimental design that aim to minimize some scalar measure of the variance of estimates of treatment effects. See, for example, Atkinson, Donev, and Tobias (2007).

Medical research has long distinguished between the statistical and clinical significance of treatment effects (e.g., Sedgwick, 2014). While the idea of clinical significance has been interpreted in various ways, many writers call an average treatment effect clinically significant if its magnitude is greater than specified value deemed minimally consequential in clinical practice. The ICH put it this way (International Conference on Harmonisation, 1999, p. 1923): "The treatment difference to be detected may be based on a judgment concerning the minimal effect which has clinical relevance in the management of patients."

Research articles reporting trial findings sometimes pose particular values of MCMDs when comparing alternative treatments for specific diseases. For example, in a study comparing drug treatments for hypertension, Materson *et al.* (1993) defined the outcome of interest to be the fraction of subjects who achieve a specified threshold for blood pressure. They took the MCMD to be the fraction 0.15, stating that this is (p. 916): "the difference specified in the study design to be clinically important." In a study evaluating a topical ointment to prevent infection after minor surgery, Sedgwick (2014) defined the outcome of interest to be the fraction of subjects who did not suffer an infection after treatment. He states (p. 1): "The smallest effect of clinical interest was an absolute decrease in the incidence of infection of 5%." c

Section 2 formalizes our general decision theoretic principles for evaluation of trial designs and treatment rules, including ε-optimality. We first present concepts and notation in the simple leading case where the decision problem is to allocate a population of observationally identical persons to two treatments. We then extend the presentation to encompass settings with multiple treatments and ones in which persons have observable covariates that may be used to differentiate their treatment.

Section 3 presents our study of trial designs that enable ε-optimal treatment. We consider trials that draw predetermined numbers of subjects at random within groups stratified by covariates and treatments. The principal analytical findings are simple sufficient conditions on sample sizes that ensure existence of ε-optimal treatment rules when outcomes are bounded. These conditions are obtained by application of Hoeffding (1963) large deviations inequalities to evaluate the performance of empirical success rules, which use the empirical distribution of the sample data to estimate the population distribution of treatment response. We also provide exact computations of minimal sample sizes enabling ε-optimality that hold when there are two treatments and outcomes are binary. In this setting, we compare our findings on trial design with those obtained using conventional practices based on the theory of hypothesis testing.

Section 4 briefly extends the analysis to situations in which the trial draws subjects from a sub-population rather than from the complete target treatment population. The concluding Section 5 sketches multiple directions for furthering the research initiated here.

2. Principles for Evaluation of Trial Designs and Treatment Rules

2.1. The Planning Problem

The setup is as in Manski (2004) and Manski and Tetenov (2007). For simplicity we initially consider the simple case with two treatments and no observable covariates.

A planner must assign one of two treatments to each member of a treatment population, denoted J. The feasible treatments are $T = \{a, b\}$. Each $j \in J$ has a response function $u_j(\cdot): T \rightarrow U$ mapping treatments $t \in T$ into individual welfare outcomes $u_j(t) \in R$. Treatment is individualistic; that is, a person's outcome may depend on the treatment he is assigned but not on the treatments assigned to others. The population is a probability space (J, Ω, P), and the probability distribution P[u(·)] of the random function u(·): T → R describes treatment response across the population. The population is "large;" formally J is uncountable and $P(j) = 0, j \in J$.

While treatment response may be heterogeneous, the members of the population are observationally identical to the planner. That is, the planner does not observe person-specific covariates that would enable systematic differentiation of treatment of different persons. However, the planner can randomly allocate persons to the two treatments with specified allocation probabilities.

A statistical treatment rule maps sample data into a treatment allocation. Let Q denote the sampling distribution generating the available data and let Ψ denote the sample space; that is, Ψ is the set of data samples that may be drawn under Q. Section 3 will focus on cases in which the data are generated in a classical trial, with Q determined by the trial design. Here we consider an abstract sampling process, encompassing both performance of trials and collection of observational data on treatment outcomes.

Let Δ denote the space of functions that map T × Ψ into the unit interval and that satisfy the adding-up conditions: $\delta \in \Delta \Rightarrow \delta(a, \psi) + \delta(b, \psi) = 1, \forall \psi \in \Psi$. Each function $\delta \in \Delta$ defines a statistical treatment rule, δ(a, ψ) and δ(b, ψ) being the fractions of the population assigned to treatments a and b when the data are ψ. This definition of an STR does not specify which persons receive each treatment, only the assignment shares. Designation of the particular persons receiving each treatment is immaterial because assignment is random, the population is large, and the planner has an additive welfare function. As δ(a, ψ) + δ(b, ψ) = 1, we use the shorthand δ(ψ) to denote the fraction assigned to treatment b. The fraction assigned to treatment a is 1 − δ(ψ).

The planner wants to maximize population welfare, which adds welfare outcomes across persons. Given data ψ, the population welfare that would be realized if the planner were to choose rule δ is

(1) $\quad U(\delta, P, \psi) \equiv E[u(a)] \cdot [1 - \delta(\psi)] + E[u(b)] \cdot \delta(\psi) \equiv \mu_a \cdot [1 - \delta(\psi)] + \mu_b \cdot \delta(\psi)$,

where $\mu_a \equiv E[u(a)] \equiv \int_J u_j(a) dP(j)$ and $\mu_b \equiv E[u(b)] \equiv \int_J u_j(b) dP(j)$ are assumed to be finite. Inspection of (1) shows that, whatever value $\psi$ may take, it is optimal to set $\delta(\psi) = 0$ if $\mu_a > \mu_b$ and $\delta(\psi) = 1$ if $\mu_a < \mu_b$. All allocations are optimal if $\mu_a = \mu_b$.

The problem of interest is treatment choice when one does not have enough knowledge of P to determine the ordering of $\mu_a$ and $\mu_b$. Hence, the planner does not know the optimal treatment.

2.2. Evaluating Treatment Rules by their State-Dependent Welfare Distributions

The starting point for development of implementable criteria for treatment choice under uncertainty is specification of a state space, say S. Thus, let $\{(P_s, Q_s), s \in S\}$ be the set of (P, Q) pairs that the planner deems possible. The planner does not know the optimal treatment if S contains at least one state such that $\mu_{sa} > \mu_{sb}$ and another such that $\mu_{sa} < \mu_{sb}$. We assume this throughout.

Considered as a function of $\psi$, $U(\delta, P_s, \psi)$ is a random variable with state-dependent sampling distribution $Q_s[U(\delta, P_s, \psi)]$. Following Wald's view of statistical decision functions as procedures, we use the vector $\{Q_s[U(\delta, P_s, \psi)], s \in S\}$ of state-dependent welfare distributions to evaluate rule $\delta$. In principle this vector is computable, whatever the state space and sampling process may be. Hence, in principle, a planner can compare the vectors of state-dependent welfare distributions yielded by different STRs and base treatment choice on this comparison.

How might a planner compare the state-dependent welfare distributions yielded by different STRs? The planner wants to maximize welfare, so it seems self-evident that he should weakly prefer rule $\delta$ to an alternative rule $\delta'$ if, in every $s \in S$, $Q_s[U(\delta, P_s, \psi)]$ equals or stochastically dominates $Q_s[U(\delta', P_s, \psi)]$. It is less obvious how one should compare rules whose state-dependent welfare distributions are not uniformly ordered in this manner, as is typically the case.

Wald evaluated statistical decision functions by their mean performance across realizations of the sampling process and this has become the standard practice in the subsequent literature. Research on treatment choice has generally adhered to this practice, an exception being the Manski and Tetenov (2014) study of quantile performance. The expected welfare yielded by rule $\delta$ in state s, denoted $W(\delta, P_s, Q_s)$, is

(2) $\quad W(\delta, P_s, Q_s) = \mu_{sa} \cdot \{1 - E_s[\delta(\psi)]\} + \mu_{sb} \cdot E_s[\delta(\psi)]$.

$E_s[\delta(\psi)] \equiv \int_\Psi \delta(\psi) dQ_s(\psi)$ is the mean (across potential samples) fraction of persons assigned treatment b.

2.3. Admissibility, Optimality, and ε-Optimality of Treatment Rules

A planner must confront the fact that the true state of nature is unknown. The concept of admissibility eliminates from consideration rules that are inferior whatever the true state may be. Rule $\delta$ is *mean inadmissible (admissible)* if there exists (does not exist) another rule $\delta'$ such that $W(\delta', P_s, Q_s) \geq W(\delta, P_s, Q_s)$ for all $s \in S$ and $W(\delta', P_s, Q_s) > W(\delta, P_s, Q_s)$ for some s.

We define rule $\delta$ to be *mean optimal* if $W(\delta, P_s, Q_s) = \max(\mu_{sa}, \mu_{sb})$ for all $s \in S$. Optimality nests admissibility: if $\delta$ is mean optimal, it necessarily is mean admissible. Mean optimality is desirable, but it is too strong to be achievable in general. The concept of mean ε-optimality relaxes mean optimality, yielding a property that may be achievable in practice.

We define rule $\delta$ to be *mean ε-optimal* for a specified $\varepsilon > 0$ if $W(\delta, P_s, Q_s) \geq \max(\mu_{sa}, \mu_{sb}) - \varepsilon$ for all $s \in S$. Section 3 shows that mean ε-optimal treatment rules exist when treatment outcomes are bounded and classical trials have sufficient finite size, whatever the state space may be. This finding makes ε-optimality a practical criterion for trial design.

Stating that an STR is ε-optimal is equivalent to stating that it has maximum regret no larger than ε. By definition, the regret of rule δ in state s is $\max(\mu_{sa}, \mu_{sb}) - W(\delta, P_s, Q_s)$. The maximum regret of δ across all states is $\max_{s \in S} [\max(\mu_{sa}, \mu_{sb}) - W(\delta, P_s, Q_s)]$. Thus, maximum regret is less than or equal to ε if and only if $\max(\mu_{sa}, \mu_{sb}) - W(\delta, P_s, Q_s) \leq \varepsilon$, all $s \in S$. It follows that mean ε-optimal STRs exist with a specified design if and only if the maximum regret of the minimax-regret rule is less than or equal to ε.

For each rule δ and state s, the smallest feasible value of $W(\delta, P_s, Q_s)$ is $\min(\mu_{sa}, \mu_{sb})$. Hence, the performance of δ in state s is necessarily consistent with ε-optimality if $|\mu_{sa} - \mu_{sb}| \leq \varepsilon$. This suggests specification of ε to equal the minimum clinically meaningful difference in the outcomes of the two treatments. The MCMD when comparing treatments a and b is the minimum value of $|\mu_a - \mu_b|$ deemed relevant to clinical practice.

2.4. Test Rules

The above discussion has placed no structure on treatment rules. It has been common to employ rules that use the outcome of a hypothesis test to choose between the two treatments. Construction of a test rule begins by partitioning the state space into disjoint subsets $S_a$ and $S_b$, where $S_a$ contains all states in which treatment a is optimal and $S_b$ contains all states in which b is optimal. Thus, $\mu_{sa} > \mu_{sb} \Rightarrow s \in S_a$, $\mu_{sa} < \mu_{sb} \Rightarrow s \in S_b$, and the states with $\mu_{sa} = \mu_{sb}$ are somehow split between the two sets. Let s* denote the unknown true state. The two hypotheses are $[s^* \in S_a]$ and $[s^* \in S_b]$.

A test rule δ partitions sample space Ψ into disjoint acceptance regions $\Psi_{\delta a}$ and $\Psi_{\delta b}$. When the data ψ lie in $\Psi_{\delta a}$, the rule accepts hypothesis $[s^* \in S_a]$ by setting $\delta(\psi) = 0$. When ψ lies in $\Psi_{\delta b}$, the rule accepts $[s^* \in S_b]$ by setting $\delta(\psi) = 1$. We use the word "accepts" rather than the traditional term "does not reject" because treatment choice is an affirmative action.

The above shows that test-based rules are *uniformly singleton*. That is, for every possible data realization, a test rule assigns the entire population to one of the two treatments. It never assigns a positive fraction of the population to each treatment. Indeed, the converse holds as well. If δ is uniformly singleton, one can collect all of the data values for which the rule assigns everyone to treatment a, call this subset of the sample space the acceptance region $\Psi_{\delta a}$, and do likewise for $\Psi_{\delta b}$. Thus, test rules and uniformly singleton rules are synonymous.

The expected welfare of a test rule has the form

$$(3) \quad W(\delta, P_s, Q_s) = \mu_{sa} \cdot Q_s[\delta(\psi) = 0] + \mu_{sb} \cdot Q_s[\delta(\psi) = 1].$$

States with $\mu_{sa} = \mu_{sb}$ are irrelevant to ε-optimality for all values of ε, so we focus on states with $\mu_{sa} \neq \mu_{sb}$. Let $R_s(\delta)$ be the state-dependent probability that δ yields an error, choosing the inferior treatment over the superior one. That is,

$$(4) \quad R_s(\delta) = Q_s[\delta(\psi) = 0] \quad \text{if } \mu_{sa} < \mu_{sb},$$
$$\phantom{(4) \quad R_s(\delta)} = Q_s[\delta(\psi) = 1] \quad \text{if } \mu_{sa} > \mu_{sb}.$$

It follows from (3) and (4) that

$$(5) \quad W(\delta, P_s, Q_s) = \min(\mu_{sa}, \mu_{sb}) \cdot R_s(\delta) + \max(\mu_{sa}, \mu_{sb}) \cdot [1 - R_s(\delta)].$$

A fundamental feature of (5) is that the state dependent error probabilities of a test rule symmetrically determine expected welfare. This contrasts with the theory of hypothesis testing, which considers tests that yield a predetermined probability of a Type I error (conventionally 0.05) and seeks a test of this type that yields an adequately small probability of a Type II error (typically 0.10 to 0.20) in some specified state.

2.5. Settings with Multiple Treatments and Observable Covariates

It is straightforward to extend the concepts and notation of Sections 2.1 through 2.3 to settings with multiple treatments and to ones in which persons have observable covariates. Let T continue to denote the set of feasible treatments. As above, each $j \in J$ has a response function $u_j(\cdot): T \to Y$ mapping treatments $t \in T$ into individual welfare outcomes $u_j(t) \in R$.

Let person j have observable covariates $x_j$ taking a value in a covariate space X; thus, $x: J \to X$ is the random variable mapping persons into their covariates. We suppose that X is finite with $P(x = \xi) > 0$, $\forall \xi \in X$. We also suppose that the covariate distribution $P(x)$ is known.

The planner can systematically differentiate persons with different observed covariates, but he cannot distinguish among persons with the same observed covariates. A feasible treatment rule is a function that assigns all persons with the same observed covariates to one treatment or, more generally, a function that randomly allocates such persons across the different treatments. Let $\Delta$ now denote the space of functions that map $T \times X \times \Psi$ into the unit interval and that satisfy the adding-up conditions: $\delta \in \Delta \implies \sum_{t \in T} \delta(t, \xi, \psi) = 1, \ \forall (\xi, \psi) \in X \times \Psi$. Then each function $\delta \in \Delta$ defines a statistical treatment rule.

With this notation, the population welfare that results if rule $\delta$ is used with sample data $\psi$ is

(6) $\quad U(\delta, P, \psi) \equiv \sum_{\xi \in X} P(x = \xi) \sum_{t \in T} \delta(t, \xi, \psi) \cdot E[u(t) | x = \xi].$

Expected welfare is

(7) $\quad W(\delta, P, Q) \equiv \int_\Psi \sum_{\xi \in X} P(x = \xi) \sum_{t \in T} \delta(t, \xi, \psi) \cdot E[u(t)|x = \xi] dQ(\psi)$

$$= \sum_{\xi \in X} P(x = \xi) \sum_{t \in T} E[\delta(t, \xi, \psi)] \cdot E[u(t)|x = \xi].$$

Here $E[\delta(t, \xi, \psi)] \equiv \int_\Psi \delta(t, \xi, \psi) dQ(\psi)$ is the expected (across potential samples) fraction of persons with covariates $\xi$ who are assigned to treatment t.

The definition of admissibility remains valid as stated in Section 2.3. To define $\varepsilon$-optimality, we need new notation for the maximum welfare achievable in each state s. This is

(8) $\quad U^*(P_s) \equiv \sum_{\xi \in X} P(x = \xi) \max_{t \in T} E_s[u(t)|x = \xi].$

Rule $\delta$ is *mean $\varepsilon$-optimal* if $W(\delta, P_s, Q_s) \geq U^*(P_s) - \varepsilon$ for all $s \in S$.

The notion of a test rule does not extend to settings with multiple treatments and/or covariates because these settings require the planner to choose among multiple actions, whereas the theory of hypothesis testing only considers a binary choice between specified null and alternative hypotheses. On the other hand, the notion of a uniformly singleton treatment rule extends immediately. Rule $\delta$ is uniformly singleton if, for each value of $(t, \xi, \psi)$, $\delta(t, \xi, \psi) = 0$ or 1. Expected welfare has the form

(9) $\quad W(\delta, P, Q) = \sum_{\xi \in X} P(x = \xi) \sum_{t \in T} Q[\delta(t, \xi, \psi) = 1] \cdot E[u(t)|x = \xi].$

## 3. Randomized Trials with Sample Sizes Enabling ε-Optimal Treatment

We now investigate the existence of ε-optimal treatment rules when the data are generated by classical randomized trials. We specifically consider trials that draw subjects at random within groups stratified by covariates and treatments. Thus, for $(t, \xi) \in T \times X$, the experimenter draws $n_{t\xi}$ subjects at random from the sub-population with covariates $\xi$ and assigns them to treatment t. The set $n_{TX} \equiv [n_{t\xi}, (t, \xi) \in T \times X]$ of stratum sample sizes defines the design. Let $n(t, \xi)$ be the realized sample of subjects with covariates $\xi$ who are assigned to treatment t. The data are the sample outcomes $\psi = [u_j, j \in n(t, \xi); (t, \xi) \in T \times X]$. We suppose throughout that the state space S contains all distributions of treatment response. Thus, the planner has no prior knowledge restricting the variation of response across treatments and covariates.

In principle, the existence of ε-optimal STRs under any design can be determined by computing the maximum regret of the minimax-regret (MMR) rule. As noted earlier, ε-optimal rules exist if and only if the MMR rule has maximum regret less than or equal to ε. In practice, determination of the MMR rule and computation of its maximum regret may be burdensome. To date, exact minimax-regret decision rules have been derived only for the case of T=2 treatments with equal or nearly-equal sample sizes. See Schlag (2006) and Stoye (2009, 2012). Hence, it is useful to have simple sufficient conditions that ensure existence of ε-optimal rules. This section provides such conditions in settings where outcomes are bounded.

### 3.1. Sufficient Conditions for ε-Optimality of Empirical Success Rules

To show that a specified trial design enables ε-optimal STRs, it suffices to consider a particular STR and to show that this rule is ε-optimal when used with this design. We focus on *empirical success*

*(ES)* rules, which use the empirical distribution of the sample data to estimate the population distribution of treatment response. Formally, let $m_{t\xi}(\psi)$ be the average outcome in treatment-covariate sub-sample $N(t, \xi)$; that is, $m_{t\xi}(\psi) \equiv (1/n_{t\xi})\sum_{j \in N(t, \xi)} u_j$. An ES rule $\delta$ assigns all persons with covariates $\xi$ to treatments that maximize $m_{t\xi}(\psi)$ over T. Thus, $\delta(t, \xi, \psi) = 0$ if $m_{t\xi}(\psi) < \max_{t' \in T} m_{t'\xi}(\psi)$.

Two analytical reasons motivate interest in ES rules when outcomes are bounded. First, it is known that these rules either exactly or approximately minimize maximum regret in various settings with two treatments when sample size is moderate (Stoye, 2009, 2012) and asymptotically (Hirano and Porter, 2009). Second, large deviations inequalities derived in Hoeffding (1963) may be used to obtain informative and easily computable upper bounds on the maximum regret of ES rules applied with any number of treatments. These upper bounds on maximum regret immediately yield sample sizes that ensure an ES rule is $\varepsilon$-optimal.

The case of two treatments has been studied previously in Manski (2004, Section 3.3), who exploited the large deviations result of Hoeffding (1963, Theorem 2) to derive an upper bound on the maximum regret of a class of ES rules that condition treatment on alternative subsets of the observable covariates of population members. The bound takes a particularly simple form when one conditions on all observable covariates and the state space includes all distributions of treatment response.

Let outcomes lie in the bounded range $[u_l, u_h]$, label the treatments $t = a$ and $t = b$, and let S index all distributions of treatment response. Manski (2004, eq. 23) showed that the maximum regret of an ES rule $\delta$ is bounded from above as follows:

(10) $\quad \max_{s \in S} U^*(P_s) - W(\delta, P_s, Q_s) \leq \tfrac{1}{2} e^{-\tfrac{1}{2}} (u_h - u_l) \sum_{\xi \in X} P(x = \xi)(n_{a\xi}^{-1} + n_{b\xi}^{-1})^{\tfrac{1}{2}}.$

Hence, an ES rule is $\varepsilon$-optimal if the trial sample sizes satisfy the inequality

(11)    $\tfrac{1}{2} e^{-\tfrac{1}{2}(u_h - u_l)} \sum_{\xi \in X} P(x = \xi)(n_{a\xi}^{-1} + n_{b\xi}^{-1})^{\tfrac{1}{2}} \leq \varepsilon$.

When the design is balanced, with $n_{t\xi} = n$ for all $(t, \xi)$, inequality (11) reduces to $(2e)^{-\tfrac{1}{2}}(u_h - u_l)n^{-\tfrac{1}{2}} \leq \varepsilon$. Hence, an ES rule with a balanced design is $\varepsilon$-optimal if $n \geq (2e)^{-1}[(u_h - u_l)/\varepsilon]^2$.

In what follows we present new findings that hold with any finite number of treatments. The cardinality of the set of treatments is denoted $|T|$. Sections 3.2 through 3.4 consider trial design when members of the population have no observable covariates. Section 3.5 extends the analysis to settings with covariates.

3.2. Large Deviation Bounds on Maximum Regret with Multiple Treatments

Propositions 1 and 2 present two alternative upper bounds on the maximum regret of an ES rule. Proposition 1 extends inequality (10) to multiple treatments while Proposition 2 uses a different large-deviations bound. We find that when the design is balanced, with $n_t = n$ for all $t$, Proposition 1 provides a tighter bound than Proposition 2 when there are two or three treatments. Proposition 2 gives a tighter bound when there are four or more treatments. Proposition 3 shows that, for any given total sample size that is an integer multiple of $|T|$, the bounds on maximum regret derived in Propositions 1 and 2 are minimized by balanced designs.

In all propositions, a design is a vector of sample sizes $(n_t, t \in T)$ and $t^* \in \operatorname{argmin}_{t \in T} n_t$ denotes a treatment with the smallest sample size. In the proofs we use the notation $\mu_{st} \equiv E_s[u(t)]$ and let $t(s)$ designate any one of the optimal treatments in state $s$; that is, $\mu_{st(s)} \geq \mu_{st}$ for all $t \in T$.

*Proposition 1*: The maximum regret of an empirical success rule $\delta$ is bounded above as follows:

(12)  $\max_{s \in S} U^*(P_s) - W(\delta, P_s, Q_s) \leq \tfrac{1}{2} e^{-\frac{1}{2}}(u_h - u_l) \sum_{t \neq t^*} (n_t^{-1} + n_{t^*}^{-1})^{\frac{1}{2}}.$

When the design is balanced, with $n_t = n$ for all t, the bound is $(2e)^{-\frac{1}{2}}(u_h - u_l)(|T| - 1)n^{-\frac{1}{2}}.$ ☐

Proof: Given that $\delta$ is an empirical success rule, $\delta(t, \psi) \leq 1[m_t \geq m_{t'}]$ for all (t, t') in T and all $\psi \in \Psi$. Therefore, $E_s[\delta(t, \psi)] \leq P_s(m_t \geq m_{t'})$ in each state s. Hence, $E_s[\delta(t, \psi)] \leq P_s(m_t \geq m_{t(s)})$. The best achievable welfare in state s is $U^*(P_s) = \mu_{st(s)}$. Hence, the regret of $\delta$ in state s is

$$U^*(P_s) - W(\delta, P_s, Q_s) = \mu_{st(s)} - \sum_{t \in T} E_s[\delta(t, \psi)]\mu_{st} = \sum_{t \neq t(s)} E_s[\delta(t, \psi)](\mu_{st(s)} - \mu_{st}) \leq$$

$$\leq \sum_{t \neq t(s)} (\mu_{st(s)} - \mu_{st}) \cdot P_s(m_t \geq m_{t(s)}).$$

Adaptation of the argument used by Manski (2004) to obtain his inequality (19) from Hoeffding's large deviations result (1963, Theorem 2) shows that

$$(\mu_{st(s)} - \mu_{st}) \cdot P(m_t \geq m_{t(s)}) \leq \tfrac{1}{2} e^{-\frac{1}{2}}(u_h - u_l)(n_t^{-1} + n_{t(s)}^{-1})^{\frac{1}{2}}.$$

It follows that

$$U^*(P_s) - W(\delta, P_s, Q_s) \leq \tfrac{1}{2} e^{-\frac{1}{2}}(u_h - u_l) \sum_{t \neq t(s)} (n_t^{-1} + n_{t(s)}^{-1})^{\frac{1}{2}}.$$

Hence, maximum regret is bounded above as follows:

$$\max_{s \in S} U^*(P_s) - W(\delta, P_s, Q_s) \leq \tfrac{1}{2} e^{-\frac{1}{2}}(u_h - u_l) \max_{s \in S} \sum_{t \neq t(s)} (n_t^{-1} + n_{t(s)}^{-1})^{\frac{1}{2}}.$$

Finally, the summation $\sum_{t \neq t(s)} (n_t^{-1} + n_{t(s)}^{-1})^{1/2}$ is maximized in a state s such that $t(s) = t^*$, where $t^*$ is a treatment with the smallest sample size. This holds because $n_{t^{**}} \geq n_{t^*}$ for any $t^{**} \neq t^*$. Hence,

$$\sum_{t \neq t^*} (n_t^{-1} + n_{t^*}^{-1})^{1/2} - \sum_{t \neq t^{**}} (n_t^{-1} + n_{t^{**}}^{-1})^{1/2} = \sum_{t \neq t^*, t^{**}} [(n_t^{-1} + n_{t^*}^{-1})^{1/2} - (n_t^{-1} + n_{t^{**}}^{-1})^{1/2}] \geq 0.$$

Thus, (12) holds.

Q. E. D.

*Proposition 2:* The maximum regret of an empirical success rule δ is bounded above by

(13) $\quad \max_{s \in S} U^*(P_s) - W(\delta, P_s, Q_s) \leq N^{-1/2}(u_h - u_l) \min_{d > 0} \dfrac{\ln\{1 + \sum_{t \neq t^*} \exp[d^2(p_t^{-1} + p_{t^*}^{-1})/8]\}}{d}$,

where $N \equiv \sum_{t \in T} n_t$ is total sample size and $p_t \equiv n_t/N$. When the design is balanced, with $n_t = n$ for all t, (13) implies the bound

(14) $\quad \max_{s \in S} U^*(P_s) - W(\delta, P_s, Q_s) \leq n^{-1/2}(u_h - u_l)(\ln T)^{1/2}.$  ☐

Proof: The proof of (13) is in four parts. Result (14) is then proved in Part V.

*I:* Fix state s and consider a treatment t(s) that is optimal in this state. Fix the sample data ψ. Let

$$D_{s[t, t(s)]}(\psi) \equiv [m_t(\psi) - m_{t(s)}(\psi)] - (\mu_{st} - \mu_{st(s)})$$

denote the amount by which $m_t(\psi) - m_{t(s)}(\psi)$ overestimates $\mu_{st} - \mu_{st(s)}$. Note that $D_{s[t(s),t(s)]}(\psi) = 0$. We first show that the welfare loss $U^*(P_s) - U(\delta, P_s, \psi)$ is bounded above by

$$U^*(P_s) - U(\delta, P_s, \psi) \leq \max_{t \in T} D_{s[t, t(s)]}(\psi).$$

To prove this inequality, let t be any treatment and observe that a necessary condition for $\delta(t, \psi) > 0$ is that $m_t(\psi) \geq m_{t(s)}(\psi)$. For any t such that $\delta(t, \psi) > 0$,

$$\mu_{st(s)} - \mu_{st} \leq [m_t(\psi) - m_{t(s)}(\psi)] - [\mu_{st} - \mu_{st(s)}] = D_{s[t, t(s)]}(\psi) \leq \max_{t' \in T} D_{s[t', t(s)]}(\psi).$$

Given that $\delta(t, \psi) \geq 0$ for all t and that $\sum_{t \in T} \delta(t, \psi) = 1$, it follows that

$$U^*(P_s) - U(\delta, P_s, \psi) = \sum_{t: \delta(t, \psi) > 0} \delta(t, \psi)(\mu_{st(s)} - \mu_{st}) \leq \max_{t \in T} D_{s[t, t(s)]}(\psi).$$

*II:* Each variable $D_{s[t, t(s)]}(\psi)$ is a sum of independent mean zero variables

$$D_{s[t, t(s)]}(\psi) = [m_t(\psi) - \mu_{st}] - [m_{t(s)}(\psi) - \mu_{st(s)}] = \sum_{j \in N(t)} (u_j - \mu_{st})/n_t - \sum_{j \in N[t(s)]} (u_j - \mu_{st(s)})/n_{t(s)}.$$

Inequality (4.16) of Hoeffding (1963) applies to each element of both sums on the right-hand side. This inequality shows that, for any $c > 0$,

$$E_s\{\exp\{c[(u - \mu_{st})/n_t)]\} \leq \exp[c^2 n_t^{-2}(u_h - u_l)^2/8]$$

for each element of the first sum and

$$E_s\{\exp\{c[(u - \mu_{st(s)})/n_{t(s)})]\} \leq \exp[c^2 n_{t(s)}^{-2}(u_h - u_l)^2/8]$$

for each element of the second sum. The statistical independence of these elements implies that

$$E_s\{\exp[c \cdot D_{s[t, t(s)]}(\psi)]\} \leq \exp[c^2(n_t^{-1} + n_{t(s)}^{-1})(u_h - u_l)^2/8].$$

*III:* The conclusion to Part I implies that the regret of δ in state s is bounded above as follows:

$$U^*(P_s) - W(\delta, P_s, Q_s) \leq E_s[\max_{t \in T} D_{s[t, t(s)]}(\psi)].$$

We use the conclusion to Part II and a proof similar to Lemma 1.3 in Lugosi (2002) to obtain an upper bound on $E_s[\max_{t \in T} D_{s[t, t(s)]}(\psi)]$. For any $c > 0$, by Jensen's inequality,

$$\exp\{c \cdot E_s[\max_{t \in T} D_{s[t, t(s)]}(\psi)]\} \leq E_s\{\exp[c \cdot \max_{t \in T} D_{s[t, t(s)]}(\psi)]\} =$$

$$= E_s\{\max_{t \in T} \exp\{c \cdot D_{s[t, ,t(s)]}(\psi)]\} \leq E_s\{\sum_{t \in T} \exp[c \cdot D_{s[t, t(s)]}(\psi)]\} =$$

$$= 1 + \sum_{t \neq t(s)} E_s\{\exp[c \cdot D_{s[t, ,t(s)]}(\psi)]\} \leq$$

$$\leq 1 + \sum_{t \neq t(s)} \exp[c^2(n_t^{-1} + n_{t(s)}^{-1})(u_h - u_l)^2/8],$$

where the last inequality follows from the conclusion to Part II. Taking the logarithm of both sides and dividing by c yields

$$E_s[\max_{t \in T} D_{s[t, t(s)]}(\psi)] \leq \frac{\ln\{1 + \sum_{t \neq t(s)} \exp[c^2(n_t^{-1} + n_{t(s)}^{-1})(u_h - u_l)^2/8]\}}{c} =$$

$$= N^{-1/2}(u_h - u_l) \; \frac{\ln\{1 + \sum_{t \neq t(s)} \exp[d^2(p_t^{-1} + p_{t(s)}^{-1})/8]\}}{d},$$

where $d = N^{-1/2}(u_h - u_l)c$.

*IV*. The conclusion to III holds in every state s. Hence, the maximum regret of $\delta$ is bounded above by

$$\max_{s \in S} U^*(P_s) - W(\delta, P_s, Q_s) \leq N^{-1/2}(u_h - u_l) \max_{s \in S} \frac{\ln\{1 + \sum_{t \neq t(s)} \exp[d^2(p_t^{-1} + p_{t(s)}^{-1})/8]\}}{d}.$$

The summation $\sum_{t \neq t(s)} \exp[d^2(p_t^{-1} + p_{t(s)}^{-1})/8]$ is maximized in a state s such that $t(s) = t^*$, where $t^*$ is a treatment with the smallest sample size. This holds because $p_{t^{**}} \geq p_{t^*}$ for any $t^{**} \neq t^*$. Hence,

$$\sum_{t \neq t^*} \exp[d^2(p_t^{-1} + p_{t^*}^{-1})/8] - \sum_{t \neq t^{**}} \exp[d^2(p_t^{-1} + p_{t^{**}}^{-1})/8] =$$
$$= \sum_{t \neq t^*, t^{**}} \{\exp[d^2(p_t^{-1} + p_{t^*}^{-1})/8] - \exp[d^2(p_t^{-1} + p_{t^{**}}^{-1})/8]\} \geq 0.$$

The above shows that

$$\max_{s \in S} U^*(P_s) - W(\delta, P_s, Q_s) \leq N^{-1/2}(u_h - u_l) \; \frac{\ln\{1 + \sum_{t \neq t^*} \exp[d^2(p_t^{-1} + p_{t^*}^{-1})/8]\}}{d}.$$

Finally, observe that the above inequality holds for all $d > 0$. This yields result (13).

*V*. If $n_t = n$ for all t, then $p_t = n/N$ for all t. It follows that

$$1 + \sum_{t \neq t^*} \exp[d^2(p_t^{-1} + p_{t^*}^{-1})/8] \leq |T| \cdot \exp[(n/N)^{-1}d^2/4].$$

Hence, (13) implies that

$$\max_{s \in S} U^*(P_s) - W(\delta, P_s, Q_s) \leq N^{-1/2}(u_h - u_l) \min_{d > 0} \frac{\ln\{|T| \cdot \exp[(n/N)^{-1}d^2/4]\}}{d}$$

$$= n^{-1/2}(u_h - u_l) \min_{e > 0} \frac{\ln[|T| \cdot \exp(e^2/4)]}{e},$$

where $e = (n/N)^{-1/2}d$. The minimum is obtained at $e = 2(\ln|T|)^{1/2}$. This implies result (14).

<div align="right">Q. E. D.</div>

*Proposition 3:* Consider any positive integer n. Among all designs with total sample size $|T| \cdot n$,

(a) bound (12) in Proposition 1 is minimized by a balanced design with $n_t = n$ for all t.

(b) bound (13) in Proposition 2 is minimized by a balanced design with $p_t = 1/|T|$ for all t.  □

Proof:

a) Bound (12) of Proposition 1 established that maximum regret is less than

$$\tfrac{1}{2}e^{-1/2}(u_h - u_l) \sum_{t \neq t^*} (n_t^{-1} + n_{t^*}^{-1})^{1/2}.$$

For a balanced design, the sum in the bound equals

$$\sum_{t \neq t^*} (n^{-1} + n^{-1})^{1/2} = (|T| - 1) 2^{1/2} n^{-1/2}.$$

For any design with $\sum_{t \in T} n_t = |T| \cdot n$, the minimum sample size is $n_{t^*} \leq n$. This and the fact that $|T| \geq 2$ imply that

(15) $\quad \sum_{t \neq t^*} (n_t + n_{t^*}) = |T| \cdot n + (|T| - 2) n_{t^*} \leq 2(|T| - 1) \cdot n.$

Applying Jensen's inequality to $f(x) = x^{-1}$, which is convex for $x > 0$, yields $n_t^{-1} + n_{t^*}^{-1} \geq 4(n_t + n_{t^*})^{-1}$. In the derivation below, we apply this inequality to the sum in bound (12), then apply Jensen's inequality to the function $f(x) = x^{-1/2}$, which is convex for $x > 0$, and then combine inequality (15) with the fact that $f(x) = x^{-1/2}$ is a decreasing function:

$$\sum_{t \neq t^*} (n_t^{-1} + n_{t^*}^{-1})^{1/2} \geq \sum_{t \neq t^*} [4(n_t + n_{t^*})^{-1}]^{1/2} = 2 \sum_{t \neq t^*} (n_t + n_{t^*})^{-1/2} \geq$$

$$\geq 2(|T| - 1) \cdot [(|T| - 1)^{-1} \sum_{t \neq t^*} (n_t + n_{t^*})]^{-1/2} \geq$$

$$\geq 2(|T| - 1) \cdot [(|T| - 1)^{-1} 2(|T| - 1) \cdot n]^{-1/2} = (|T| - 1) 2^{1/2} n^{-1/2}.$$

This shows that the bound for any design with total sample size $|T| \cdot n$ is no smaller than the bound with a balanced design.

(b) Bound (13) of Proposition 2 established that maximum regret is less than

$$N^{-1/2}(u_h - u_l) \min_{d > 0} \frac{\ln\{1 + \sum_{t \neq t^*} \exp[d^2(p_t^{-1} + p_{t^*}^{-1})/8]\}}{d}.$$

For a balanced design and any $d > 0$, the sum in the bound equals

$$\sum_{t \neq t^*} \exp[d^2(p_t^{-1} + p_{t^*}^{-1})/8] = \sum_{t \neq t^*} \exp\{d^2[(1/|T|)^{-1} + (1/|T|)^{-1}]/8\} = (|T| - 1)\exp[d^2|T|/4].$$

We will show that for any $(p_t, t \in T)$ such that $\sum_{t \in T} p_t = 1$,

$$\sum_{t \neq t^*} \exp[d^2(p_t^{-1} + p_{t^*}^{-1})/8] \geq (|T| - 1) \exp[d^2|T|/4].$$

This result, which holds for all $d > 0$, and the fact that $\ln(\cdot)$ is an increasing function show that the bound for any design with total sample size $|T| \cdot n$ is no smaller than the bound with a balanced design.

Applying Jensen's inequality to $f(x) = x^{-1}$, which is convex for $x > 0$, yields $p_t^{-1} + p_{t^*}^{-1} \geq 4(p_t + p_{t^*})^{-1}$. Given that $\exp(\cdot)$ is increasing, it follows that

(16) $\quad \sum_{t \neq t^*} \exp[d^2(p_t^{-1} + p_{t^*}^{-1})/8] \geq \sum_{t \neq t^*} \exp[(d^2/2) \cdot (p_t + p_{t^*})^{-1}].$

Applying Jensen's inequality to the convex function $f(x) = \exp(x)$ yields

(17) $\quad \sum_{t \neq t^*} \exp[(d^2/2) \cdot (p_t + p_{t^*})^{-1}] \geq (|T| - 1) \exp\{(|T| - 1)^{-1} \sum_{t \neq t^*}[(d^2/2) \cdot (p_t + p_{t^*})^{-1}]\} =$

$= (|T| - 1) \exp[(d^2/2) \cdot (|T| - 1)^{-1} \sum_{t \neq t^*} (p_t + p_{t^*})^{-1}].$

Applying Jensen's inequality to $f(x) = x^{-1}$, which is convex for $x > 0$, yields

$$(|T| - 1)^{-1} \sum_{t \neq t^*} (p_t + p_{t^*})^{-1} \geq [(|T| - 1)^{-1} \sum_{t \neq t^*} (p_t + p_{t^*})]^{-1} = \{(|T| - 1)^{-1}[1 + (|T| - 2)p_{t^*}]\}^{-1}.$$

Given that $|T| - 2 \geq 0$ and $p_{t^*} \leq 1/|T|$, it follows that $1 + (|T| - 2)p_{t^*} \leq 2(|T| - 1)/|T|$. Given that $f(x) = x^{-1}$ is a decreasing function, it follows that

(18)  $(|T| - 1)^{-1} \sum_{t \neq t^*} (p_t + p_{t^*})^{-1} \geq \{(|T| - 1)^{-1} [2(|T| - 1)/|T|]\}^{-1} = |T|/2.$

Combining (16), (17), and (18) with the monotonicity of $\exp(\cdot)$ yields

$$\sum_{t \neq t^*} \exp[d^2(p_t^{-1} + p_{t^*}^{-1})/8] \geq (|T| - 1)\exp[d^2|T|/4].$$

Q. E. D.

Table 1 presents the values of bounds (12), (13), and (14) for balanced designs. The three bounds vary identically with $(u_h - u_l)n^{-\frac{1}{2}}$ but differently with the number of treatments. Proposition 1 provides a better bound for $|T| \leq 3$, while Proposition 2 provides a better bound for $|T| \geq 4$. Bound (14) of Proposition 2 is simpler to compute than bound (13) and is only marginally larger.

We have also computed bounds (12) and (13) for various unbalanced designs. We again find that bound (12) is better for $|T| \leq 3$ and bound (13) is better for $|T| \geq 4$. These results are not shown in the table.

Table 1: Bounds in Propositions 1 and 2 for Balanced Designs with n Subjects per Treatment

| $|T| =$ | 2 | 3 | 4 | 5 | 6 | 7 | |
|---|---|---|---|---|---|---|---|
| Bound (12) | 0.4289 | 0.8578 | 1.2866 | 1.7155 | 2.1444 | 2.5733 | $\cdot (u_h - u_l)n^{-\frac{1}{2}}$ |
| Bound (13) | 0.6539 | 0.9279 | 1.0892 | 1.1999 | 1.2827 | 1.3481 | $\cdot (u_h - u_l)n^{-\frac{1}{2}}$ |
| Bound (14) | 0.8326 | 1.0481 | 1.1774 | 1.2686 | 1.3386 | 1.3950 | $\cdot (u_h - u_l)n^{-\frac{1}{2}}$ |

3.3. Implications of the Bounds for ε-Optimality of Empirical Success Rules

Propositions 1 and 2 imply sufficient conditions on sample sizes for ε-optimality of ES rules. If the upper bound on maximum regret with a specified trial design is less than or equal to ε, then ES rules are ε-optimal with this design.

The findings are particularly simple with balanced designs. Then bound (12) of Proposition 1 implies that an ES rule is ε-optimal if $n \geq (2e)^{-1}(|T| - 1)^2[(u_h - u_l)/\varepsilon]^2$. Bound (14) of Proposition 2 implies that an ES rule is ε-optimal if $n \geq \ln|T| \cdot [(u_h - u_l)/\varepsilon]^2$. Table 1 gives the threshold sample size for bound (13) of Proposition 2 for $|T| \leq 7$, which is $[(u_h - u_l)/\varepsilon]^2$ times the square of the relevant constant shown in the table.

To illustrate the findings, consider the Materson *et al.* (1993) study of treatment for hypertension described in the Introduction. The outcome is binary with $u_l = 0$ and $u_h = 1$. The study compared seven drug treatments and specified 0.15 as the MCMD. We cannot know how the authors of the study, who reported results of traditional hypothesis tests, would have specified ε had they sought to achieve ε-optimality. If they were to set ε = 0.15, application of bound (13) shows that an ES rule is ε-optimal if the number of subjects per treatment arm is at least $(1.3481)^2 \cdot (0.15)^{-2} = 80.8$. The actual study has an approximately balanced design, with between 178 and 188 subjects in each treatment arm.

It is important to bear in mind that Propositions 1 and 2 only imply simple sufficient conditions on sample sizes for ε-optimality of ES rules, not necessary ones. Proposition 1, for example, could be sharpened for balanced designs by replacing Hoeffding's inequality by the result of Bentkus (2004, Theorem 1.2)[2] and further improvements should be possible. In general it is difficult to compute the exact maximum regret of ES rules, hence difficult to determine how conservative the propositions are. An exception occurs when there are two treatments and outcomes are binary. Then the maximum regret of various decision rules can be computed numerically without large deviations bounds. The next section computes necessary and sufficient sample sizes for ε-optimality of an ES rule and rules using conventional hypothesis tests.

---

[2] The Bentkus inequality is expressed in terms of a tail probability of a binomial distribution and the resulting regret bound has to be evaluated numerically for each n. For large values of n, the regret bound could be up to 23.5% smaller than (12).

3.4. Exact Computations for Binary Outcomes, Two Treatments, and Balanced Designs

Consider a setting with binary outcomes, two treatments, and a balanced design. Table 2 provides exact numerical computations of the minimum sample size that enables ε-optimality for three treatment rules, for various values of ε. We compute the maximum regret of each treatment rule over all feasible states $(\mu_{sa}, \mu_{sb}) \in [0,1] \times [0,1]$ for n = 1, 2, 3 , . . . and report the smallest sample size enabling ε-optimality. Outcomes being binary, $u_l = 0$ and $u_h = 1$.

The first column shows the minimum sample size (per treatment arm) that yields ε-optimality if an ES rule is used. The ES rule used for this table assigns the status quo treatment a to ½ of the population if the number of successes is the same in both treatment groups. This is the minimax-regret rule for binary outcomes (Stoye, 2009). The second and third columns display the minimum sample sizes that yield ε-optimality of rules based on classical one-sided hypothesis tests. There is no consensus on what hypothesis test should be used to compare two proportions. We report results based on the widely used one-sided two-sample z-test, which is based on an asymptotic normal approximation. The second column sets the probability of a Type I error at 0.05 and the third column at 0.01.

The findings are remarkable. A sample as small as 1 observation per treatment arm makes the ES rule ε-optimal when ε = 0.15 and a sample of size 145 suffices when ε = 0.01. The minimum sample sizes required for ε-optimality of the test rules are orders of magnitude larger. If the z-test is used with a 0.05 probability of Type I error, a sample of size 16 is required when ε = 0.15 and 3488 when ε = 0.01. If a 0.01 probability of a Type I error is used, the sample sizes have to be more than double these values.

The final column of the table reports the sufficient sample sizes for ε-optimality of the ES rule obtained using bound (12) of Proposition 1. Comparison of this column and the first one quantifies the conservatism of bound (12), which does not use the information that outcomes are binary and which relies on the Hoeffding large-deviations inequality for bounded outcomes. We find that the sufficient sample sizes provided by the bound are roughly ten times the size of the exact minimum sample sizes, depending on the value of ε. This strongly suggests that it is worthwhile to compute exact minimum sample sizes whenever it is tractable to do so.

Table 2: Minimum Sample Sizes Per Treatment Enabling ε-Optimal Treatment Choice

| MCMD | ES Rule binary outcomes | One-Sided 5% z-Test binary outcomes | One-Sided 1% z-Test binary outcomes | Proposition 1 Bound bounded outcomes |
|---|---|---|---|---|
| ε = 0.01 | 145 | 3488 | 7963 | 1840 |
| ε = 0.03 | 17 | 382 | 879 | 205 |
| ε = 0.05 | 6 | 138 | 310 | 74 |
| ε = 0.10 | 2 | 33 | 79 | 19 |
| ε = 0.15 | 1 | 16 | 35 | 9 |

It has been traditional to use hypothesis tests to choose a sample size that achieves specified statistical power rather than ε-optimality. We now consider the one-sided z-tests from this perspective. Equation (19) gives the sample size (per arm) required to achieve specified probabilities (α, β) of Type I and Type II errors when ($\mu_{sa}$, $\mu_{sb}$) take specified values, obtained with a commonly used asymptotic normal approximation (e.g., Fleiss, 1973):

$$(19) \quad n_{power} = \frac{\{z_\alpha[2(\mu^*(1-\mu^*))]^{1/2} + z_\beta[\mu_{sa}(1-\mu_{sa}) + \mu_{sb}(1-\mu_{sb})]^{1/2}\}^2}{(\mu_{sb}-\mu_{sa})^2},$$

29where $\mu^* = \frac{1}{2}(\mu_{sa} + \mu_{sb})$, $z_\alpha = \Phi^{-1}(1 - \alpha)$, $z_\beta = \Phi^{-1}(1 - \beta)$, and $\Phi$ is the standard normal distribution function. Among states of nature that have a given effect size $\Delta = \mu_{sb} - \mu_{sa}$, the largest sample size is necessary in the state where $\mu_{sa} = (1 - \Delta)/2$ and $\mu_{sb} = (1 + \Delta)/2$. Then (19) simplifies to

$$(20) \quad n_{power} = \frac{\{z_\alpha + z_\beta(1 - \Delta^2)^{1/2}\}^2}{2\Delta^2}.$$

Table 3 shows the sample sizes determined by applying (20) for various values of $\Delta$, with $\alpha = 0.05$ and $\beta = 0.20$ or $0.10$. The table also shows the exact maximum regret that results when this trial design is combined with subsequent application of the z-test treatment rule with $\alpha = 0.05$. The table reports findings for the same values of $\Delta$ as we considered for $\varepsilon$ in Table 2. The effect size $\Delta$ that a conventional trial designer uses to define an alternative hypothesis of interest may or may not be the same as the value of $\varepsilon$ that a planner would use when seeking $\varepsilon$-optimality.

Table 3: Minimum Sample Sizes Achieving Specified Statistical Power

| $\Delta$ | $n_{power}$ ($\alpha=.05$, $\beta=.20$) | Maximum Regret of the z-test rule | $n_{power}$ ($\alpha = .05$, $\beta = .10$) | Maximum Regret of the z-test rule |
|---|---|---|---|---|
| $\Delta = 0.01$ | 30912 | 0.0034 | 42818 | 0.0029 |
| $\Delta = 0.03$ | 3434 | 0.0102 | 4756 | 0.0086 |
| $\Delta = 0.05$ | 1236 | 0.0167 | 1711 | 0.0144 |
| $\Delta = 0.10$ | 309 | 0.0338 | 427 | 0.0291 |
| $\Delta = 0.15$ | 137 | 0.0501 | 189 | 0.0417 |

Observe that knowledge of the statistical power of a test rule does not suffice to determine its maximum regret. The reason is that power is not evaluated at the parameter value where maximum regret is achieved. For example, when the experiment is designed to yield the error probabilities ($\alpha = 0.05$, $\beta = 0.20$) for the effect size $\Delta = 0.10$, maximum regret at values of ($\mu_{sa}$, $\mu_{sb}$) such that $\mu_{sb} - \mu_{sa} = 0.10$ is approximately $(0.10) \cdot (0.20) = 0.02$ by design. Maximum regret across all states of nature, however, occurs at states such that $\mu_{sb} - \mu_{sa} \neq 0.10$ and turns out to equal 0.0338. For the design with $\beta = 0.10$ and $\Delta = 0.10$, maximum regret in states such that $\Delta = 0.1$ is approximately $(0.10) \cdot (0.10) = 0.01$, but maximum regret across all states is 0.0291. Overall maximum regret tends to be attained in states with smaller than the specified effect size, which have higher probabilities of Type II error.

3.5 ε-Optimality of Empirical Success Rules with Observable Covariates

The above analysis has assumed that members of the population have no observable covariates that may be used to condition treatment choice. Suppose now that persons have observable covariates taking values in a finite set X and that the planner can execute a trial with (treatment, covariate)-specific sample sizes [$n_{t\xi}$, (t, ξ) ∈ T × X]. We consider the ES rule defined in Section 3.1, which assigns all persons with covariates ξ to treatments that maximize $m_{t\xi}(\psi)$ over T, $m_{t\xi}(\psi)$ being the average outcome in sub-sample N(t, ξ).

There are at least two reasonable ways that a planner may wish to evaluate ε-optimality in this setting. First, he may want to achieve ε-optimality within each covariate group. This interpretation requires no new analysis. The planner should simply define each covariate group to be a separate population of interest and then apply the analysis of Sections 3.2 through 3.4 to each group. The design that achieves group-specific ε-optimality with minimum total sample size equalizes sample sizes across groups.

Alternatively, the planner may want to achieve ε-optimality within the overall population, without requiring that it be achieved within each covariate group. This is the interpretation given in Section 2.5, when we defined rule $\delta$ to be $\varepsilon$-optimal if $W(\delta, P_s, Q_s) \geq U^*(P_s) - \varepsilon$ for all $s \in S$. In this case, the design that achieves $\varepsilon$-optimality with minimum total sample size does not equalize sample sizes across groups.

Propositions 1 and 2 easily extend to provide sample sizes sufficiently large to yield the latter interpretation of $\varepsilon$-optimality. Applying Proposition 1 to each group and aggregating the bounds across groups implies that the maximum regret of ES rules is bounded above by

$$(12') \quad \max_{s \in S} U^*(P_s) - W(\delta, P_s, Q_s) \leq \tfrac{1}{2} e^{-\tfrac{1}{2}} (u_h - u_l) \sum_{\xi \in X} P(x = \xi) \sum_{t \neq t^*(\xi)} (n_{t\xi}^{-1} + n_{t^*(\xi)\xi}^{-1})^{\tfrac{1}{2}},$$

where $t^*(\xi) \in \operatorname{argmin}_{t \in T} n_{t\xi}$ denotes a treatment with the smallest sample size among individuals with covariate value $\xi$. When the design is balanced across treatments for each covariate, with $n_{t\xi} = n_\xi$ for all t, the bound is

$$(21) \quad (2e)^{-\tfrac{1}{2}} (|T| - 1) (u_h - u_l) \sum_{\xi \in X} P(x = \xi) n_\xi^{-\tfrac{1}{2}}.$$

The analogous extension of Proposition 2 yields

$$(13') \quad \max_{s \in S} U^*(P_s) - W(\delta, P_s, Q_s) \leq$$

$$\leq (u_h - u_l) \sum_{\xi \in X} P(x = \xi) N_\xi^{-\tfrac{1}{2}} \min_{d > 0} \frac{\ln\{1 + \sum_{t \neq t^*(\xi)} \exp[d^2(p_{t\xi}^{-1} + p_{t^*(\xi)\xi}^{-1})/8]\}}{d},$$

where $N_\xi \equiv \sum_{t \in T} n_{t\xi}$ is total sample size for individuals with covariate value $\xi$ and $p_{t\xi} \equiv n_{t\xi}/N_\xi$. When the design is balanced across treatments for each covariate, $N_\xi^{-\tfrac{1}{2}} = |T|^{-\tfrac{1}{2}} n_\xi^{-\tfrac{1}{2}}$, $p_{t\xi} = 1/|T|$ for all $(t, \xi)$, and the bound in (13') simplifies to

$$(22) \quad |T|^{-\frac{1}{2}} \min_{d > 0} \frac{\ln\{1 + (|T| - 1) \exp[d^2|T|/4]\}}{d} (u_h - u_l) \sum_{\xi \in X} P(x = \xi) \cdot n_\xi^{-\frac{1}{2}}.$$

Bounds (21) and (22) can easily be evaluated for any candidate treatment-balanced design to verify whether it suffices to enable ε-optimal treatment rules. The constants preceding $\sum_{\xi \in X} P(x = \xi) \cdot n_\xi^{-\frac{1}{2}}$ in these bounds are given in Table 1 for $|T| \leq 7$.

Given a predetermined maximum total sample size N, minimizing bounds (21) and (22) is achieved by choosing $(n_\xi, \xi \in X)$ to minimize $\sum_{\xi \in X} P(x = \xi) \cdot n_\xi^{-\frac{1}{2}}$ subject to the constraint $\sum_{\xi \in X} n_\xi \leq N/|T|$. Given that the objective function is decreasing in each $n_\xi$, the constraint binds. The Lagrangian expression of the constrained minimization problem is

$$(23) \quad L[(n_\xi, \xi \in X), \lambda] \equiv \sum_{\xi \in X} P(x = \xi) \cdot n_\xi^{-\frac{1}{2}} + \lambda(\sum_{\xi \in X} n_\xi - N/|T|).$$

A simple approximation to the minimization problem results if one treats $(n_\xi, \xi \in X)$ as continuous variables rather than as integer sample sizes. Then the first order conditions for minimization of $L(\cdot, \cdot)$ yield

$$(24) \quad -\tfrac{1}{2} P(x = \xi) \cdot n_\xi^{-3/2} + \lambda = 0, \text{ all } \xi \in X.$$

This implies that $n_\xi = (2\lambda)^{-\frac{2}{3}} P(x = \xi)^{\frac{2}{3}}$. It follows that, to solve problem (23), the relative sample sizes for any pair $(\xi, \xi')$ of covariate values have the approximate ratio

$$(25) \quad n_\xi/n_{\xi'} = [P(x = \xi)/P(x = \xi')]^{\frac{2}{3}}.$$

For the case when the covariate takes two values, a similar result is obtained by Schlag (2006).

A planner who uses (25) to choose the trial design makes the covariate-specific sample size increase with the prevalence of the covariate group in the population, albeit less than proportionately. Covariate-specific maximum regret commensurately decreases with the prevalence of the covariate group.

4. ε-Optimality of Empirical Success Rules with Partial External Validity

We mentioned at the outset that medical conventions for choosing sample size in clinical trials pertain to classical trials possessing perfect internal and external validity. However, practical trials usually have only partial validity. Hence, experimental data may only partially identify the mean treatment response vector $\{E[u(t)|x = \xi], t \in T, \xi \in X\}$ in the target treatment population.

The concept of ε-optimality readily extends to such situations. To illustrate we consider here the common case in which the experimental sample is representative only of a part of the target treatment population. This is common because experimental subjects generally are persons who meet specified criteria and who consent to participate in the trial.

Let $z_j = 1$ if a person from the target population is a member of the sub-population randomly sampled in an experiment and $z_j = 0$ if a person is not in the sampling frame. Denote the fraction of persons who are not in the sampling frame by $\kappa \equiv P_s(z_j = 0)$. Assume for simplicity that $\kappa$ is known and constant across all states $s \in S$. We will consider the case when persons have no observed covariates. We use the notation $\mu_{st|z=1} \equiv E_s[u(t)|z = 1]$ for mean treatment response of persons in the sampling frame of the experiment and $\mu_{st|z=0} \equiv E_s[u(t)|z = 0]$ for mean response in the unsampled subpopulation. We impose no restrictions on the state space S, allowing for any relationship between $P[u(t)|z = 0]$ and $P[u(t)|z = 1]$. This implies that the experimental data obtained from the sampled subpopulation reveal nothing about treatment response in the unsampled subpopulation.

The maximum welfare achievable in state s equals

$$U^*(P_s) \equiv \max_{t \in T} E_s[u(t)] = \max_{t \in T} \{(1 - \kappa)\mu_{st|z=1} + \kappa\mu_{st|z=0}\},$$

whereas the welfare achieved by treatment rule $\delta$ is

$$W(\delta, P_s, Q_s) \equiv \sum_{t \in T} E_s[\delta(t, \psi)] \cdot \{(1 - \kappa)\mu_{st|z=1} + \kappa\mu_{st|z=0}\}.$$

The regret of rule $\delta$ is bounded above by

$$U^*(P_s) - W(\delta, P_s, Q_s) \leq (1 - \kappa) \sum_{t \in T} E_s[\delta(t, \psi)] \cdot \{\max_{t' \in T} \mu_{st'|z=1} - \mu_{st|z=1}\}$$

$$+ \kappa \sum_{t \in T} E_s[\delta(t, \psi)] \cdot \{\max_{t' \in T} \mu_{st'|z=0} - \mu_{st|z=0}\}$$

(26) $$\leq (1 - \kappa) \sum_{t \in T} E_s[\delta(t, \psi)] \cdot \{\max_{t' \in T} \mu_{st'|z=1} - \mu_{st|z=1}\} + \kappa(u_h - u_l).$$

The first inequality holds because $\max_{t \in T}\{(1 - \kappa)\mu_{st|z=1} + \kappa\mu_{st|z=0}\} \leq (1 - \kappa) \max_{t \in T} \mu_{st|z=1} + \kappa \max_{t \in T} \mu_{st|z=0}$. The second holds because $\max_{t' \in T} \mu_{st'|z=1} - \mu_{st|z=1} \leq (u_h - u_l)$.

Let $\delta$ be an empirical success rule that uses the experimental data to estimate mean treatment response in the sampled sub-population and then applies the findings to choose treatments in the complete target population. The results of Propositions 1 and 2 can be applied to the first term of (26) by simply replacing $\mu_{st}$ by $\mu_{st|z=1}$ in the proofs. Doing so yields the following bounds on the maximum regret of empirical success rules:

(27)     $\max_{s \in S} U^*(P_s) - W(\delta, P_s, Q_s) \le (1 - \kappa) \cdot \tfrac{1}{2} e^{-1/2}(u_h - u_l) \sum_{t \ne t^*} (n_t^{-1} + n_{t^*}^{-1})^{1/2} + \kappa(u_h - u_l),$

(28)     $\max_{s \in S} U^*(P_s) - W(\delta, P_s, Q_s) \le (1 - \kappa) N^{-1/2}(u_h - u_l) \min_{d > 0} \dfrac{\ln\{1 + \sum_{t \ne t^*} \exp[d^2(p_t^{-1} + p_{t^*}^{-1})/8]\}}{d}$

$+ \kappa(u_h - u_l).$

Both bounds are minimized by choosing a balanced sample with $n_t = n$ for each treatment arm. With balanced samples, bound (27) is lower than bound (28) for $|T| < 4$. A necessary condition to guarantee ε-optimality using these bounds is that $\kappa(u_h - u_l) \le \varepsilon$. If this condition is satisfied, calculation of sufficient sample size based on bounds (27) and (28) is the same as it was for bounds (12) and (13), except that ε is replaced by $[\varepsilon - \kappa(u_h - u_l)]/(1 - \kappa)$.

5. Conclusion

Choosing sample sizes in clinical trials to enable ε-optimal treatment rules would align trial design directly with the objective of informing treatment choice. In contrast, the conventional practice of choosing sample size to achieve specified statistical power in hypothesis testing is only loosely related to treatment choice. We share with Bayesian statisticians who have written on trial design the objective of informing treatment choice. We differ in our application of the frequentist statistical decision theory developed by Wald. In particular, we have observed that ε-optimality is equivalent to having maximum regret no larger than ε.

There are numerous potentially fruitful directions for further research of the type initiated here.

Among them, we will mention consideration of alternative sampling processes and treatment rules as well as alternative assumptions regarding outcomes, the population, and the state space.

Our analysis in Section 3 considered trials in which the design draws a predetermined number of subjects from each covariate group and assigns a predetermined number of them to each treatment. An alternative class of designs specifies a probability distribution for drawing subjects and assigning them to treatments. With such a design, the numbers of subjects who have particular covariates and receive specific treatment are ex ante random rather than predetermined. Study of probabilistic designs is more complex than is study of designs with predetermined subject allocations. Nevertheless, the analysis of this paper remains useful as an "inner loop" that conditions on realized sample sizes. Manski (2004, Section 3.4) proved an extension of Proposition 1 to probabilistic designs when there are two treatments. The derivation performed there is applicable more generally.

Our analysis focused on ES treatment rules, with secondary attention to rules using z-tests. ES rules are natural to consider. They are familiar, computationally simple, and well-behaved statistically. Nevertheless, the most desirable treatment rule from the perspective of ε-optimality obviously is the minimax-regret rule. Given any trial design, this rule by definition yields ε-optimality for the smallest possible value of ε. The minimax-regret rule generally does not have a closed form expression and must be studied numerically. Development of improved methods for computation of the rule is an important subject for further research.

Two further technical questions for future research are analysis of ε-optimality when outcomes are unbounded and when the population is finite. Our large-deviations analysis of the ES rule assumed that outcomes are bounded and our exact computations of maximum regret maintained the stronger assumption that outcomes are binary. This paper has maintained the analytically convenient idealization that the trial draws subjects from a large population formalized as a continuum of persons. It would be valuable to weaken these assumptions.

Finally, we call attention to the fact that our analysis in Section 3 supposed that the state space contains all distributions of treatment response. Thus, the planner was assumed to have no prior knowledge restricting the variation of response across treatments and covariates. This assumption, which has been traditional in the study of clinical trials, is highly advantageous in the sense that it yields generally applicable findings. Nevertheless, it is unduly conservative in circumstances where some credible knowledge of treatment response is available. One may, for example, think it credible to maintain some assumptions on the degree to which treatment response may vary across treatments or covariate groups. The implications of such assumptions for minimax-regret treatment have been explored in Stoye (2012). When such assumptions are warranted, it may be valuable to impose them. As the state space shrinks, the minimum sample needed to achieve $\varepsilon$-optimality logically cannot increase and may decrease in size. The open research question is to characterize how various assumptions on treatment response affect the achievability of $\varepsilon$-optimality.

39Materson, B., D. Reda,, W. Cushman, B. Massie, E. Freis, M. Kochar, R. Hamburger, C. Fye, R. Lakshman, J. Gottdiener, E. Ramirez, and W. Henderson (1993), "Single-Drug Therapy for Hypertension in Men: A Comparison of Six Antihypertensive Agents with Placebo," *The New England Journal of Medicine*, 328, 914-921.

Materson, B. and D. Reda (1994), "Correction: Single-Drug Therapy for Hypertension in Men," *The New England Journal of Medicine*, 330, 1689.

Materson, B., D. Reda, and W. Cushman (1995). "Department of Veterans Affairs Single-Drug Therapy of Hypertension Study: Revised Figures and New Data," *American Journal of Hypertension*, 8, 189-192.

Schlag, K. (2006), "Eleven – Tests Needed for a Recommendation," *EUI Working Papers,* ECO No. 2006/2. URI: http://hdl.handle.net/1814/3937.

Sedgwick, P. (2014), "Clinical Significance versus Statistical Significance," *BMJ*, 348:g2130. www.bmj.com/content/348/bmj.g2130, accessed October 11, 2014.

Spiegelhalter, D. (2004), "Incorporating Bayesian Ideas into Health-Care Evaluation," *Statistical Science*, 19, 156-174.

Spiegelhalter, D., L. Freedman, and M. Parmar (1994), "Bayesian Approaches to Randomized Trials" (with discussion), *Journal of the Royal Statistical Society Series A*, 157, 357-416.

Stoye J. (2009), "Minimax Regret Treatment Choice with Finite Samples," *Journal of Econometrics,* 151, 70-81.

Stoye, J. (2012), "Minimax Regret Treatment Choice with Covariates or with Limited Validity of Experiments," *Journal of Econometrics*, 166, 157-165.

Tetenov, A. (2012), "Statistical Treatment Choice Based on Asymmetric Minimax Regret Criteria," *Journal of Econometrics*, 166, 157-165.

Wald A. (1950), *Statistical Decision Functions,* New York: Wiley.